\documentclass[floatfix, aip, jcp]{revtex4-1}

\usepackage[utf8]{inputenc}

\usepackage{graphicx}
\usepackage{dcolumn}
\usepackage{bm}
\usepackage{hyperref}
\usepackage{url}

\usepackage{xcolor}

\usepackage[T1]{fontenc}
\usepackage{mathptmx}
\usepackage{etoolbox}

\usepackage[version=4]{mhchem}

\makeatletter
\def\@email#1#2{%
 \endgroup
 \patchcmd{\titleblock@produce}
  {\frontmatter@RRAPformat}
  {\frontmatter@RRAPformat{\produce@RRAP{*#1\href{mailto:#2}{#2}}}\frontmatter@RRAPformat}
  {}{}
}%
\makeatother

\newcommand{\nc}[0]{n_{\mathrm{c}}}
\newcommand{\invivo}{in vivo}
\newcommand{\invitro}[0]{in vitro}

\newcommand{\kon}[0]{k_{+}}
\newcommand{\koff}[0]{k_{\textrm{off}}}

\begin{document}

\title[]{A universal phase-plane model for in vivo protein aggregation}
\author{Matthew W. Cotton}
 \affiliation{%
Yusuf Hamied Department of Chemistry, University of Cambridge, UK
}%
 \affiliation{UK Dementia Research Institute, Cambridge, UK}
\affiliation{ 
Mathematical Institute, University of Oxford, UK%
}%
\author{Alain Goriely}%
\affiliation{ 
Mathematical Institute, University of Oxford, UK%
}%

\author{David Klenerman}
 \affiliation{%
Yusuf Hamied Department of Chemistry, University of Cambridge, UK%
}%
 \affiliation{UK Dementia Research Institute, Cambridge, UK}
\author{Georg Meisl$^*,$}
 \email{gm373@cam.ac.uk}
 \affiliation{%
Yusuf Hamied Department of Chemistry, University of Cambridge, UK%
}%
 \affiliation{UK Dementia Research Institute, Cambridge, UK}

\date{\today}

\begin{abstract}

Neurodegenerative diseases are driven by the accumulation of protein aggregates in the brain of affected individuals. The aggregation behaviour \invitro{} is well understood and driven by the equilibration of a super-saturated protein solution to its aggregated equilibrium state. However, the situation is altered fundamentally in living systems where active processes consume energy to remove aggregates. It remains unclear how and why cells transition from a state with predominantly monomeric protein, which is stable over decades, to one dominated by aggregates.  Here, we develop a simple but  universal theoretical framework to describe cellular systems that include both aggregate formation and removal. Using a two-dimensional phase-plane representation, we show that the interplay of aggregate formation and removal generates cell-level bistability, with a bifurcation structure that explains both the emergence of disease and the effects of therapeutic interventions. We explore a wide range of aggregate formation and removal mechanisms and show that phenomena such as seeding arise robustly when a minimal set of requirements on the mechanism are satisfied. By connecting in vitro aggregation mechanisms to changes in cell state, our framework provides a general conceptual link between molecular-level therapeutic interventions and their impact on disease progression.
\end{abstract}

\maketitle

\section{Background}

Protein aggregates are known to play a key role in the onset of a variety of neurodegenerative diseases (NDDs)~\cite{chiti_protein_2006,chiti_protein_2017}. The physical mechanisms governing the formation of aggregates from monomeric proteins have been well studied using \invitro{} experiments. Theoretical descriptions based on chemical kinetics provide precise predictions of the rate at which the protein monomers are converted into the aggregates. In recent decades, it has been demonstrated that these models show good agreement with experimental data across a wide range of proteins~\cite{ferrone_kinetics_1985, Knowles2009,  meisl_molecular_2016, meisl_uncovering_2022}. Fitting these models can determine reaction rates for the different mechanisms, and which processes dominate the conversion from the monomeric to aggregated state. However, while the same processes likely still take place in living systems, these well-established models are generally insufficient for describing aggregation in living systems: They assume the initial state contains a supersaturated solution of monomeric protein, and the kinetics is governed by the equilibration into the aggregated state. To understand the transition to a disease state \invivo{},  it is essential to also account for active processes present in living systems. In particular, clearance mechanisms that remove aggregates can help maintain a metastable state where toxic aggregates are removed at the same rate at which they are produced. 

Such models are essential for understanding the effect of several therapeutic interventions currently in development: Promising monoclonal antibody therapies, such as lecanemab~\cite{van_dyck_lecanemab_2023}, donanemab~\cite{sims_donanemab_2023}, and aducanumab~\cite{vaz_role_2022} aim to support the body's  removal mechanisms by targetting aggregates of the A$\beta$ peptide. Other therapeutic approaches, such as antisense therapies~\cite{rinaldi_antisense_2018, cole_-synuclein_2021, devos_tau_2017, mummery_tau-targeting_2023}, or other epigenetic editor technologies~\cite{neumann_brainwide_2024}, aim to reduce the protein concentrations and thus the aggregation potential of the system. To predict quantitatively the effect of these interventions requires robust models of aggregation and removal in living systems.

\begin{figure}
    \centering
    \includegraphics[width=0.9\columnwidth]{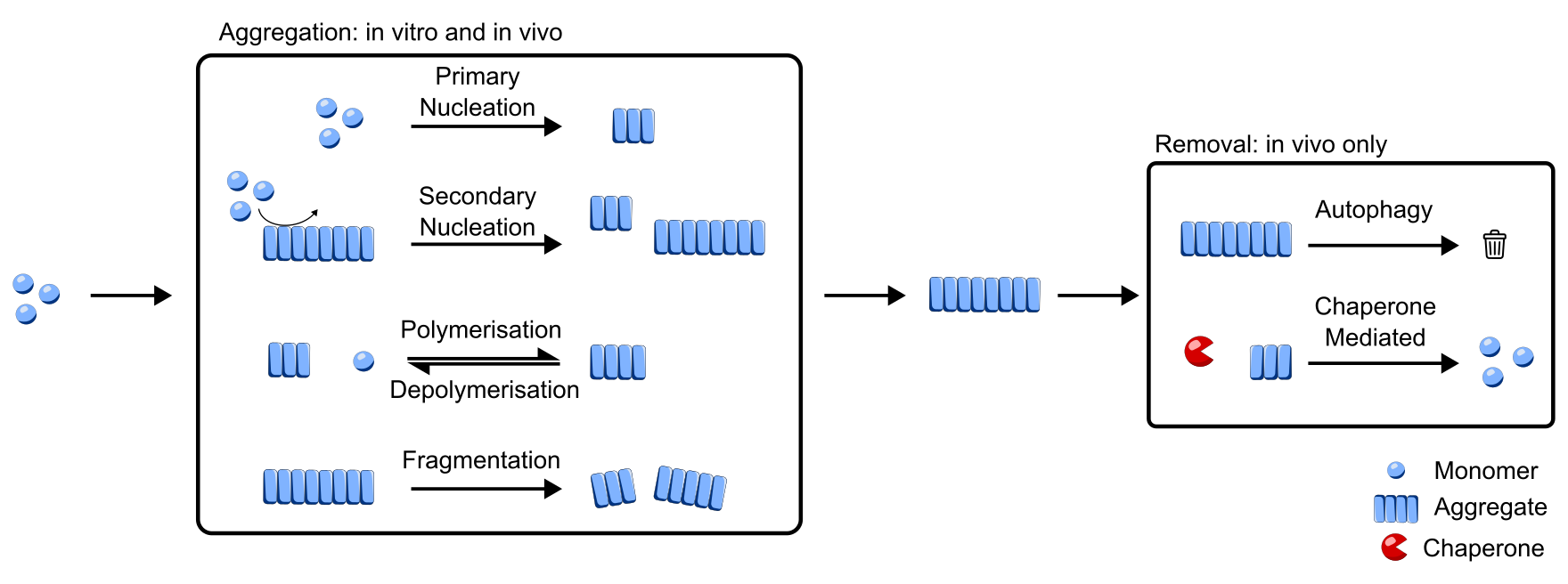}
    \caption{Multiple reactions can convert monomeric protein into an aggregated state. Additionally, in living cells aggregates are removed from cells by active processes.}
    \label{fig:scheme}
\end{figure}

In this work, we develop a general, theoretical framework to describe the aggregation state of cells, imposing a minimal set of conditions and determining the robust general features of behaviour. 
The binary state of systems, \textit{healthy} or \textit{aggregating}, naturally emerges from this description and we study how transitions between these states take place when disease initiates or when cures are administered. We generalise our earlier work in which we studied the stability of a specific minimal model of aggregate removal~\cite{thompson_role_2021}(see section \ref{subsec:constantclearance}), and introduce the two-dimensional phase-plane framework as the central tool for understanding system behaviour. We uncovered universal features of these transition by showing that the basic bifurcation structure is preserved across a wide range of scenarios, with minimal constraints on the mechanisms. This bifurcation structure provides a single explanatory framework for  widely observed phenomena, including: seeding, aggregates in healthy cells, and disappearance of aggregates upon administration of monomer-lowering therapies~\cite{miller_tau_2021, andrewsLargescaleVisualisationAsynuclein2024, cole_-synuclein_2021}.

\section{\invitro{} models of aggregation}

We assume that the intrinsic mechanisms of aggregate formation observed \invitro{} also drive the \invivo{} aggregation. It will form the basis of our model, so we briefly overview these here~\cite{Knowles2009, cohen_nucleated_2011}. The rates of these processes may be significantly different in living systems compared to test tubes, but there is significant evidence across different proteins that \invitro{} mechanisms are predictive of \invivo{} behaviour~\cite{meisl_uncovering_2022}.

New aggregates are created by \textit{primary nucleation}, \textit{secondary nucleation} and \textit{fragmentation}. Primary nucleation spontaneously converts the $\nc$ monomers into an aggregate of length of $\nc$ with rate constant $k_n$. Secondary nucleation is an autocatalytic conversion of $n_2$ monomers into an aggregate of length of $n_2$, occurring on the surface of existing aggregates with a rate constant $k_2$. Fragmentation is the splitting of a longer aggregate into two smaller aggregates which we assume occurs with equal probability for any monomer-monomer bonds with rate $k_{-}$. Additionally, aggregates change size via the addition or removal of a protein monomer on the end of an aggregate, which we call \textit{elongation} and \textit{depolymerisation} and which occur at rates $\kon$ and $\koff$ respectively. We expect that aggregates below the nucleus size $\nc$  are unstable and thus neglect them in calculations of total aggregate mass. We therefore also require $n_2\geq\nc$. The reverse reactions of nucleation and fragmentation have a negligible impact on the kinetics of aggregate formation and are thus neglected, as is commonly done~\cite{cohen_nucleated_2011}. Note that in most practical applications, i.e. when experimental data are fitted, the parameters $n_c$ and $n_2$ become effective reaction orders of a coarse-grained multi-step process and can no longer be easily interpreted directly as the sizes of the nucleated species, as detailed in~\cite{meisl_molecular_2016, Meisl2017a}.

Defining the concentration of monomers in the system as $m$ and the concentration of aggregates of length $i$ at time $t$ as $f(t, i)$, we use mass action kinetics to determine the rates of the different aggregation processes. Combining all of these processes gives the master equation~\cite{Knowles2009} that describes the evolution of the population of aggregates of a given length $i$, 
\begin{equation}
\begin{split}
    \frac{\text{d}f(t, i)}{\text{d}t} &= \delta_{i, n_C}k_n m^{n_C} + \delta_{i, n_2} k_2 m^{n_2} \left(\sum_{j=\nc}^{\infty} j f(t, j)\right) \\
    &+ 2\kon m (f(t, i-1)-f(t, i)) + 2k_\text{off} (f(t, i+1)-f(t, i)) \\
    &- k_{-}(i-1)f(t, i) + 2k_{-}\left(\sum_{j=i+1}^{\infty}f(t, j)\right).
    \end{split}
    \label{eq:pi_evolution}
\end{equation}
In general, this infinite system of coupled ordinary differential equations is difficult to solve, however with suitable assumptions, they often can be approximated by a closed system of equations for the evolution of the moments of the length distribution~\cite{Knowles2009, cohen_nucleated_2011}. The $\eta^{\text{th}}$ moment of the aggregate length distribution is given by $Q^{(\eta)} = \sum_{i=\nc}^\infty i^\eta f(t, i)$. Two particularly important moments, that are most commonly measured experimentally, are the $0^{\text{th}}$ and $1^{\text{st}}$ moments of the distribution. These are the total number concentration of aggregates, $P=\sum f(t, i)$ and the total aggregate mass concentration, $M=\sum i f(t, i)$,  respectively. Assuming that  fragmentation and depolymerisation are negligible ($k_{-}=\koff=0$), the evolution of $M$ and $P$ is then given by a closed system for $M$ and $P$:
\begin{align}
    \frac{\text{d}M}{\text{d}t} &= 2 \kon m P + \nc k_n m^{\nc}+ n_2 k_2 m^{n_2} M, \label{eq:Mvitro} \\
    \frac{\text{d}P}{\text{d}t} &= k_n m^{n_c}+k_2 m^{n_2} M. \label{eq:Pvitro}
\end{align}

\section{Extending aggregation kinetics to \invivo{} environments}

While \invitro{} the total protein concentration remains constant, in living systems, there are additional processes in the `life cycle' of a protein: specifically how proteins are produced and removed. Many proteins that aggregate during disease also fulfil functional roles and thus will likely be subject to  active homeostasis mechanisms that maintain their concentration. For example, the microtubule-associated protein tau can stabilise microtubules and regulate transport by molecular motors~\cite{alberts_molecular_2008}, but its aggregation causes tauopathies such as Alzheimer's Disease. Generally, the differential equation for monomer can be written as
\begin{equation}
    \frac{\text{d}m}{\text{d}t} = \epsilon^{-1}\left( \gamma - \lambda_1 m \right) + \text{aggregation kinetics},
    \label{eq:minvivo}
\end{equation}
where $\gamma$ and $\lambda_1$ are scaled rate constants for production and removal of the monomer respectively and $\epsilon$ is a parameter that separates the timescales of monomer production and removal and the timescales of aggregation. There are two limiting cases to consider in modelling the kinetics of aggregation \invivo: (1) the total protein concentration is constant and gets depleted by conversion into aggregates, as in the \invitro{}  case (large $\epsilon$), and (2) the free monomeric protein concentration is constant, $m(t)=m_0$, and is not depleted (small $\epsilon$).  We consider initially the latter case of a constant concentration of free monomeric protein before generalizing it in later sections. In order to maintain the monomer concentration precisely, the cellular production and removal need to be much faster than the aggregation kinetics and thus $\epsilon$ is small.  The leading order solution to (\ref{eq:minvivo}) is $m=\gamma/\lambda_1 = m_0$. In the subsequent analysis we only consider this leading order behaviour. The reality is likely somewhere in between the two limiting cases where the timescales are not as well separated~\cite{weiKineticModelsReveal2024}, however the leading order behaviour is sufficient to capture key features of the disease.

Protein synthesis can thus be addressed relatively easily. However, a more significant effect on the aggregation behaviour is the fact that in living systems aggregates can also be removed via active processes in the cells or surrounding tissue, for example by autophagy or by expulsion form the cell~\cite{morimoto_cell-nonautonomous_2020}. This introduces a new term to the master equation:
\begin{equation}
\begin{split}
    \frac{\text{d}f(t, i)}{\text{d}t} &= \delta_{i, n_C}k_n m^{n_C} + \delta_{i, n_2} k_2 m^{n_2} \left(\sum_{j=\nc}^{\infty} j f(t, j)\right) \\
    &+ 2\kon m (f(t, i-1)-f(t, i)) + 2k_\text{off} (f(t, i+1)-f(t, i)) \\
    &- k_{-}(i-1)f(t, i) + 2k_{-}\left(\sum_{j=i+1}^{\infty}f(t, j)\right)\\
    &- \lambda_i,
    \end{split}
    \label{eq:pi_cleared}
\end{equation}
where $\lambda_i$ is the removal rate for aggregates of each size. The functional form of the removal could depend on the concentration of aggregates at each length, the monomer concentration, or other parameters, i.e. $\lambda_i=\lambda(i, m, f(t,j))$, however we write it as $\lambda_i$ here for simplicity.

The exact mechanisms of aggregate removal and their kinetics remain unknown in many living systems, and we thus investigate broad classes of mechanisms to establish robust and common behaviours. We make a set of minimal assumptions that will apply throughout this work, namely that the rate of the removal mechanism is independent of the monomer concentration and that the fibril growth rate is proportional to the monomer concentration. We first consider the simplest removal mechanism, to showcase the different regimes of cell behaviour and derive the key determinants of the system. In particular, we establish the  behaviour as a function of both monomer and aggregate concentration as a characteristic "fingerprint" of the system. We then consider increasingly complex descriptions of removal and aggregation, and use these monomer-aggregate phase planes to demonstrate the common features across mechanisms. The monomer-aggregate phase planes provide a universal way to describe transitions to disease and provide a readily interpretable framework to guide the development of future therapeutics.

\subsection{Unbound removal describes a switch between stable and runaway aggregation} \label{subsec:constantclearance}

We begin by considering the simplest removal kinetics: the removal rate is proportional to the number of aggregates and independent of size so that $\lambda_i=\lambda \times f(t, i)$, and so the removal rate theoretically has no upper bound, see also Thompson \textit{et al.}~\cite{thompson_role_2021}.

When the removal is proportional to the number of aggregates, the moment equations define a linear system, since $m=m_0$ is also constant. We define $\textbf{q}=(P, M)^\text{T}$ and the evolution of the system is
\begin{equation}
    \dot{\textbf{q}} =
    \underbrace{\left(
    \begin{array}{cc}
    -\lambda & k_2 m_0^{n_2} \\
    2 \kon m_0 & n_2 k_2 m_0^{n_2}-\lambda  \\
    \end{array}
    \right)}
    _{\textbf{A}}
    \textbf{q}
    +
    \underbrace{\left(
    \begin{array}{c}
    k_n m_0^{\nc} \\
    \nc k_n m_0^{\nc} \\
    \end{array}
    \right)}
    _{\textbf{b}},
    \label{eq:momentEvomatform}
\end{equation}
where the dot indicates the time derivative and the matrix $\textbf{A}$ and vector ${\textbf{b}}$ are defined in the equation. Since generically $\textbf{A}$ is not singular, we can write the solution of this linear system in compact notation as
\begin{equation}
    \textbf{q} = -\textbf{A}^{-1}\textbf{b}+e^{\textbf{A}t}\left(\textbf{q}_0+\textbf{A}^{-1}\textbf{b}\right),
    \label{eq:momentSolvematform}
\end{equation}
where $\textbf{q}_0$ is a vector of the initial aggregate number and aggregate mass concentration. The steady state behaviour is determined by the eigenvalues of $\textbf{A}$, which are
\begin{equation}
    \nu_{\pm} = \frac{1}{2} \left(k_2 m_0^{n_2} n_2 \pm \sqrt{8 \kon m_0 k_2 m_0^{n_2} + \left(n_2 k_2 m_0^{n_2}\right)^2} - 2 \lambda \right).
    \label{eq:A_eig}
\end{equation}
The system always has $\nu_{-} \leq 0$, however the sign of $\nu_+$ depends on the rate parameters and the removal constant, $\lambda$. If $\nu_+>0$, then the system has no steady state and both the mass and the number of aggregates increase exponentially. However, for $\nu_+<0$  there exists a steady state solution for $t\rightarrow\infty$, $\textbf{q}^* = -\textbf{A}^{-1}\textbf{b}$.

At this steady state we can also find the average aggregate length, $\bar{l}^*$, by taking the ratio of mass and number concentrations, this is
\begin{equation}
    \bar{l}^* = \frac{M^*}{P^*} = \frac{2 \kon m_0 + \nc \lambda}{\lambda+(\nc-n_2)k_2 m_0^{n_2}}.
    \label{eq:lbarConstClear}
\end{equation}
The stability and steady-state length of the system is independent of the primary nucleation rate. However the primary nucleation rate does affect the steady state aggregate mass, $M^*$,
\begin{equation}
    M^* = \frac{k_n m_0^{\nc}(2 \kon m_0 + \nc \lambda)}{\lambda^2 - k_2 m_0^{n_2}(2 \kon m_0 + n_2\lambda)}.
\end{equation}

Given a set of rate constants we define the \textit{critical removal}, $\lambda^{\text{(crit)}}$, that determines whether or not the aggregate mass is finite as $t\to\infty$. Solving (\ref{eq:A_eig}) for $\nu_+=0$ gives
\begin{equation}
    \lambda^{\text{(crit)}} = \frac{1}{2} \left(k_2 m_0^{n_2} n_2 + \sqrt{8 \kon k_2 m_0^{n_2+1} + \left(n_2 k_2 m_0^{n_2}\right)^2}\right).
    \label{eq:2-critclear}
\end{equation}
For $\lambda>\lambda^{\text{(crit)}}$ the system approaches a finite \textit{steady state} and when $\lambda<\lambda^{\text{(crit)}}$, we have \textit{runaway aggregation} where the mass of aggregates grows exponentially as $M(t)\approx M(t=0)e^{\nu_+ t}$.

However, to understand how a system switches between these two states, a more useful interpretation is to fix the removal rate and determine the maximum monomer concentration that leads to a steady state solution. The rate of aggregate formation increases as the monomer concentration increases, whereas the removal rate is unaffected. Thus we expect a transition from a system that has a steady-state solution to one that undergoes runaway aggregation as the monomer concentration increases. The critical stability condition is
\begin{equation}
    -2 k_2 \kon m_0^{n_2+1} - n_2 k_2 \lambda m_0^{n_2} +\lambda^2 = 0
    \label{eq:2-critmon}
\end{equation}
which has exactly one positive solution that defines the critical monomer concentration $m_0^{\text{(crit)}}$. At monomer concentrations above $m_0^{\text{(crit)}}$ the system undergoes runaway aggregation and below $m_0^{\text{(crit)}}$ the aggregate mass approaches a steady state. Viewing the transition in terms of the monomer concentration is a useful new perspective as the monomer concentration is an alternative accessible therapeutic target compared to the removal rate, and corresponds to new treatment technologies, such as antisense oligonucleotides, that alter the expressed monomer protein $m_0$~\cite{cole_-synuclein_2021, mummery_tau-targeting_2023}. 

\begin{figure}
    \centering
    \includegraphics[width=0.8\columnwidth]{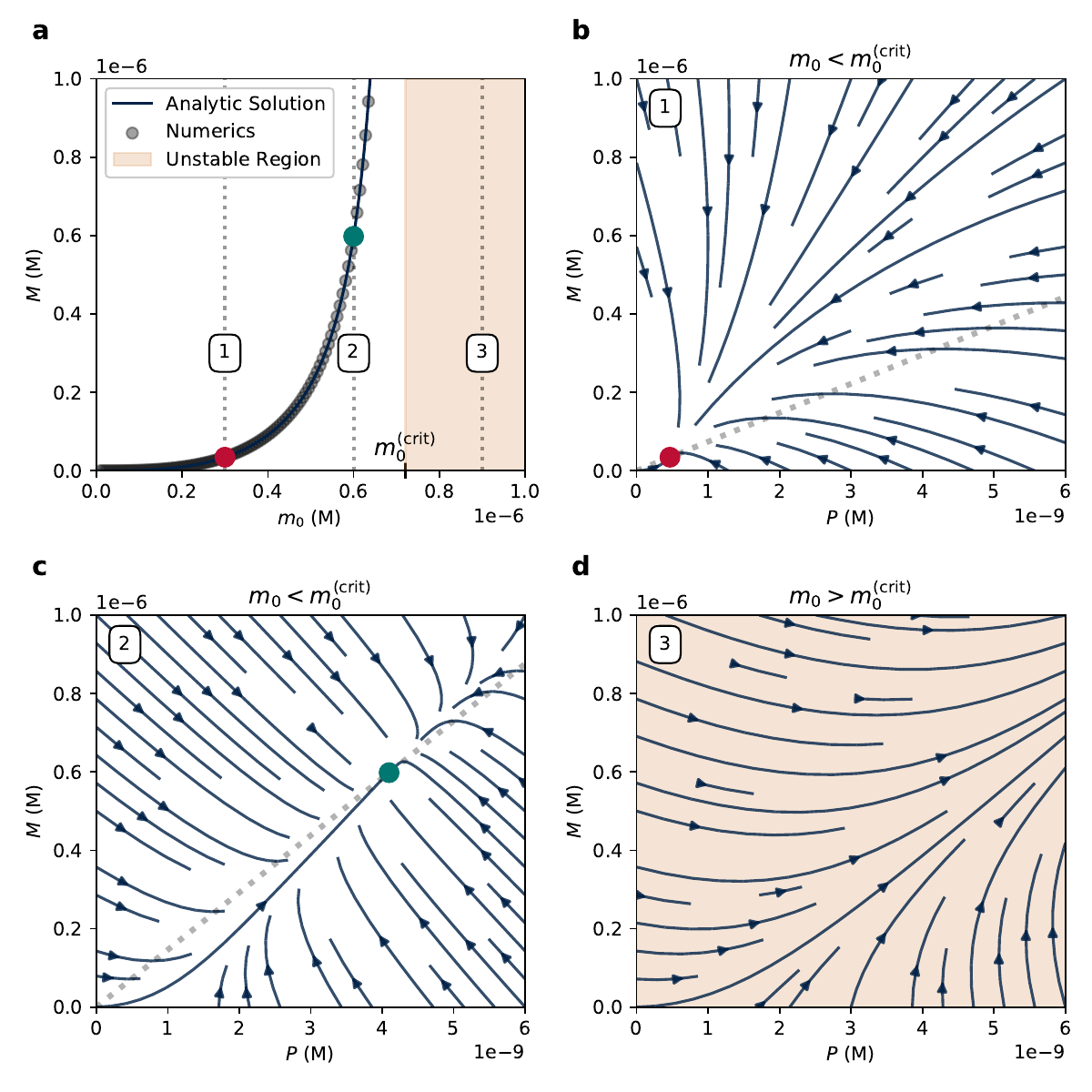}
    \caption{Exact steady states and dynamics for the unbound removal model. (a) The steady state aggregate mass diverges as the monomer concentration approaches the critical value, $m_0^{\text{(crit)}}\approx7.21\times10^{-7}\text{M}$, and for larger monomer concentrations the system undergoes runaway aggregation. The steady state aggregate mass from the numeric scheme agrees well with exact solution. (b)-(d) Show the flow of $M$ and $P$, given by (\ref{eq:momentEvomatform}), at different monomer concentrations. When $m_0<m_0^{\text{(crit)}}$ the flows approach a steady state, shown in (b) with $m_0=3\times10^{-7}\text{M}$ and (c) $6\times10^{-7}\text{M}$. When $m_0>m_0^{\text{(crit)}}$, both $M$ and $P$ grow exponentially, as shown in (d) with $9\times10^{-7}\text{M}$. System parameters: $k_n=4800.0\text{M}^{-1}\text{hr}^{-1}$, $\kon= 1.2\times10^{8}\text{M}^{-1}\text{hr}^{-1}$, $n_2=2$, $\nc=2$, $\lambda=1.0\text{hr}^{-1}$, $k_2=1.1\times10^{10}\text{M}^{-2}\text{hr}^{-1}$. Simulation parameters: $\text{d}t=0.001\text{hr}$, $T=1500.0\text{hr}$, $N=10000$.}
    \label{fig:steadyConstClear}
\end{figure}

The top left panel of Fig.~\ref{fig:steadyConstClear} shows the steady state aggregate mass concentration as a function of the monomer concentration. At $m_{0}^{\text{(crit)}}$, the steady state aggregate mass diverges and the shaded region shows the \textit{pathological state}, in which the aggregates will proliferate exponentially. In addition to identifying which cell systems we expect to be healthy or in disease, this representation provides a clear visualisation into how transitions between these two cell states take place. It is also a useful representation to generalise and compare different models and we will use it to do so extensively in the remainder of this work (see Section \ref{sec:phaseplanestructure}).
The predicted steady state aggregate mass was verified by numerical simulations, where we evolved a population of aggregates with a finite maximum length, according to \eqref{eq:pi_cleared} (see Appendix \ref{app:numdeets}). In the remaining panels (b)-(c), we show the flow of the first two moments of the length distribution as they approach equilibrium for three specific values of $m_0$. In particular, we note that when a steady state exists, (b) and (c), the fixed point is a global attractor. When $m_0>m_0^{\text{(crit)}}$ there is no positive fixed point, and the long time behaviour of $\mathbf{q}$ is determined by the eigenvector of $\mathbf{A}$ corresponding to the positive eigenvalue $\nu_+$. The ratio of the components of this eigenvector gives the average length for high aggregate mass, which is
\begin{equation}
    \bar{l} = \frac{m_0^{-\frac{n_2}{2}} \sqrt{k_2 n_2^2 m_0^{n_2}+8 \kon m_0}+\sqrt{k_2} n_2}{2 \sqrt{k_2}}.
    \label{eq:lbar-growth}
\end{equation}
On average fibrils are many hundreds to thousands of monomers long, so we expect the rate of growth of existing fibrils to be significantly faster than the nucleation of new fibrils, i.e. $k_2 m_0^{n_2} \ll \kon m_0$ and thus $\bar{l} \approx \sqrt{2\kon m_0/k_2m_0^{n_2}}$~\cite{cohen_nucleated_2011-2}. When $n_2=2$ the flow converges onto the line $M/P=\sqrt{2\kon/k_2m_0}$ and Fig.~\ref{fig:steadyConstClear}(d) begins to show this behaviour at high aggregate concentrations.

\subsection{A maximum cellular removal capacity reveals a cell-level bistability} \label{subsec:boundedclearance}

The proportional removal described above gives rise to an analytically solvable model that provides insight into the role of removal in neurodegenerative diseases. However, it is limited in its ability to describe the reality of living systems, because it assumes removal is infinitely efficient. As the aggregate mass increases, the proportional removal rate would therefore also increase without bound. However, active removal requires both energy and the availability of specific biomolecular machinery, for example to expel aggregates from a cell or to break the monomer-monomer bonds. Both the energy resources and molecules of the clearance machinery available to a cell are finite, imposing a fundamental limit on the maximum removal rate. Furthermore, aggregates may also induce toxic effects on the cell, creating feedback that impairs the efficiency of removal mechanisms~\cite{ahern_modelling_2025}. We build on the model developed in Section \ref{subsec:constantclearance} by considering more realistic kinetics and show how these naturally impose an upper limit on the removal rate.

The specific functional form governing the kinetics of aggregate removal determines how the rate approaches its upper limit. For now, we assume that the removal process follows enzyme-like kinetics, and investigate generalisations of this mechanism in later sections. As an example, in the ubiquitin (Ub)-proteasome system, aggregates are tagged with Ub, unfolded, and subsequently cleaved by the proteasome. Similarly, in chaperone-mediated autophagy, chaperones deliver aggregates to lysosomes for degradation~\cite{ciechanover_degradation_2015, barral_roles_2004}. Both processes can be modelled as involving an essential removal component (E), such as Ub or a chaperone, that binds to aggregates, forms a removable complex, and is then recycled back into the system.

For an aggregate $A_i$, of length $i$ and a removal component $E$, the removal reaction can be modelled as a typical catalytic reaction
\begin{equation}
        \ce{A_i + E <=>[k_i^b][k_i^d] C_i ->[k^{c}_i] E + Removal Products}
    \label{eq:MM}
\end{equation}
where $k_i^b$ and $k_i^d$ are the binding and dissociation rates of the aggregate and removal components and $k^{c}_i$ is the rate constant describing the removal of the aggregate. The concentration of $A_i$ is given by $f(t, i)$ and the concentration of $E$ is $f_e(t)$. The removal component binds to an aggregate of size $i$ to form a complex $C_i$, with concentration $c_i(t)$, which can either dissociate or be broken down and removed.

This setup is similar to the Michaelis–Menten (MM) description for reactions kinetics where the $E$ component is the enzyme and the aggregate is the substrate. However, we now have a series of different sized aggregates that can in principle all compete for the same removal component. We can calculate the expected rates in the system in the same way as the single substrate analysis: assume that the total $E$ component in the system is constant, $f_e^{T} = f_e(t) + \sum_i c_i(t)$, and that the aggregate binding is at equilibrium for all lengths, $k_i^b f_e(t) f(t, i) = k_i^d c_i(t)$. From these expressions, we obtain
\begin{equation}
    c_i(t) = f_e^{T}\frac{k_i^b}{k_i^d}\left(\frac{f(t, i)}{1+\sum_j f(t, j) \frac{k_j^b}{k_j^d}}\right).
\end{equation}
Factors that reduce the systems ability to remove aggregates, such as ageing, can easily be modelled by a decrease in $f_e^{T}$. The removal of aggregates of size $i$ in the master equation is $\lambda_i = k_i^{c}c_i(t)$. In general this functional form  will prevent us from obtaining a closed pair of moment equations. However if we assume that the rates are independent of aggregate size, $k_i^{c}=k^{c}$, $k_i^{b}=k^{b}$ and $k_i^{d}=k^{d}$, then the master equations reduce to the typical MM kinetics, as all aggregates act effectively as the same single substrate species (generalisations of this, for example taking into account increased probability of binding for larger aggregates, are discussed below). Following this assumption, we obtain the following system of moment equations with modified removal:
\begin{align}
    \frac{\text{d}M}{\text{d}t}&= \nc k_n m_0^{\nc} + n_2 k_2 m_0^{n_2} M + 2 \kon m_0 P - \frac{\lambda M}{1+P/K_{\lambda, P}}
    \label{eq:dM_MM_P} \\
    \frac{\text{d}P}{\text{d}t}&= k_n m_0^{\nc} + k_2 m_0^{n_2} M - \frac{\lambda P}{1+P/K_{\lambda, P}},
    \label{eq:dP_MM_P}
\end{align}
where we have defined $\lambda = k^{c} f_e^T(t)k^b/k^d$ and $K_{\lambda, P} = k^d/k^b$. At low aggregate concentrations, when $P \ll K_{\lambda, P}$, the removal is proportional to the concentration as before, with an equivalent number removal rate $\approx \lambda P$ as in the proportional removal case. However for high aggregate concentrations $P \gg K_{\lambda, P}$, the removal saturates with a maximum number removal rate $\lambda K_{\lambda, P}$.

\begin{figure}
    \centering
    \includegraphics[width=0.8\columnwidth]{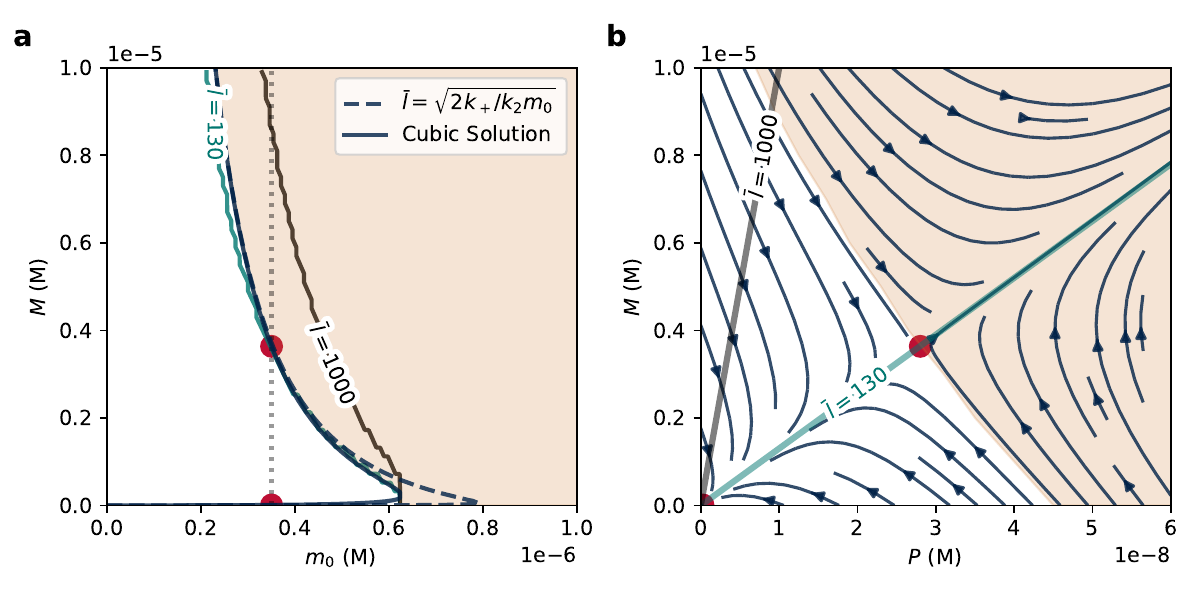}
    \caption{Exact steady states and dynamics for the MM-like removal model. (a) Steady state values of $M$ are plotted for both the full cubic system given by (\ref{eq:dM_MM_P}) and (\ref{eq:dP_MM_P}) as well as the reduced model with $\bar{l}=\sqrt{2k_{+} / k_2 m_0}$, (\ref{eq:reducedMM}). The shaded region indicates the unstable region of the reduced model and the stability boundary from numerics are shown for two different values of $\bar{l}$ of the seed. (b) The flow of $M$ and $P$, given by (\ref{eq:dM_MM_P}) and (\ref{eq:dP_MM_P}), with $3.5\times10^{-7}\text{M}$. Additional system parameters are: $k_n=5000.0\text{M}^{-1}\text{hr}^{-1}$, $\kon= 1.2\times10^{8}\text{M}^{-1}\text{hr}^{-1}$, $n_2=2$, $\nc=2$, $\lambda=2.5\text{hr}^{-1}$, $K_{\lambda, P}=1.0\times10^{-8}\text{M}$, $k_2=4.0\times10^{10}\text{M}^{-2}\text{hr}^{-1}$. Simulation parameters: $dt=0.001\text{hr}$, $T=100.0\text{hr}$ $N=10000$.}
    \label{fig:steadyMM}
\end{figure}

Solving for the steady state of the system, $\text{d}M/\text{d}t=\text{d}P/\text{d}t=0$, gives three sets of solutions for $(M^*,P^*)$. For typical system parameters at low monomer concentration, two of these sets have real positive solutions for both $M^*$ and $P^*$. However, as the monomer concentration increases, it reaches a critical monomer concentration where these two solutions undergo a saddle-node bifurcation and merge, such that for monomer concentrations above this critical value there no longer exists a positive steady state, as shown in Fig.~\ref{fig:steadyMM}(a). This parallels the behaviour described above for proportional removal with no maximum rate: when the monomer concentration is large we expect the mass of aggregates in a system to exhibit runaway growth. However, below this critical monomer concentration the introduction of a maximum removal rate alters the behaviour significantly. There are now two solutions for $(M^*,P^*)$ with the lower one corresponding to an attractive fixed point and the upper one to a repulsive one . Panel (b) of Fig.~\ref{fig:steadyMM} plots the flow of $M$ and $P$ and shows the basin of attraction of the fixed point. Systems that are below the critical monomer concentration, but outside the basin of attraction will still display runaway aggregation. Physically, this corresponds to a system in which the aggregate load is so high that the limited removal rate cannot keep up, despite the monomer concentration being below its critical value.

In practice, this means that even at a low monomer concentration, below the critical concentration, the introduction of preformed aggregates can result in runaway aggregation. This of course corresponds to the widely observed \textit{seeding} phenomenon, where cells are exposed to preformed aggregates, triggering the transition into the pathological runaway aggregation state~\cite{miller_tau_2021, tuck_cholesterol_2022, cottonNeurodegenerationEmergesCellular2025}. 

The above description also allows us to understand how the size distribution of these preformed fibrils affects whether or not a system will transition into the unstable region. For example, the introduction of a few long aggregates might not trigger runaway aggregation whereas an equivalent mass of many short aggregates would. Aggregates of different lengths are being removed at the same rate, but the same mass of smaller aggregates induce faster fibril growth, due to more growing ends. If removal processes display a different length bias, it will be reflected in the length dependence of this seeding effect. The shape of the stable region in the $M-P$ plot (Fig.~\ref{fig:steadyMM}b) highlights this fact.
This length dependence is thus useful both to understand seeding experiments \invivo{} and to investigate the properties of the removal process. It is important to not only control the total mass of aggregates added but also to control the length distribution of aggregates that will be used to seed a system to investigate the \invivo{} dynamics. A different length distribution, but the same aggregate mass, can in fact alter the fate of identical cells and so conclusions drawn from comparing the aggregation kinetics with uncontrolled length distributions may be misleading.

The length dependence of seeding may also have consequences in the spreading of pathology within the brain. The prion-like hypothesis suggests that pathological aggregates can be transported (either actively or passively) into neighbouring brain regions, seeding these regions and causing the formation of new aggregates, leading for example to \textit{Braak staging} in Parkinson's Disease~\cite{braak_staging_2003}. The general connectivity between different regions of the brain has been explored as driving the spread of pathology~\cite{putra_braiding_2021, brennan_role_2023}. Given the length-dependent seeding suggested here, a better understanding of the different transport properties of different aggregate sizes will enable a more accurate modelling of the spread of the disease through these connections. For example, short aggregates might be expected to diffuse faster and thus be transported between regions more quickly. This length-dependent transport can significantly alter the rate at which the disease spreads, or exaggerate the spreading effects of certain transport mechanisms. This is a similar mode of action to dimeric enzymes that can experience enhanced reactivity by dissociating into monomers, diffusing at a faster rate, and dimerise again to carry out function~\cite{agudo-canalejo_cooperatively_2020}.

\subsection{Monomer-aggregate phase plane captures key features of aggregation dynamics}

The long-time dynamics of aggregates determines whether a cell is in a low aggregate steady state, which we here call \textit{healthy}, or in a state of runaway aggregation, which we here also refer to as \textit{pathological}. Thus, understanding whether specific parameter values lead to stable or runaway aggregation is key to understanding disease dynamics and in particular the onset and incidence of disease. To provide an easily interpretable overview of the possible system behaviours, we project the full system dynamics onto a two dimensional phase plane defined by the monomer and aggregate concentrations, as detailed in the following.

When the aggregate removal and other rates are prescribed, the system and its dynamics are determined by three variables, $M$, $P$ and $m_0$. In the current limit, we model $m_0$ as unaffected by the aggregation kinetics, however we keep the monomer concentration as a system variable so that this framework can be applied to study monomer reducing therapies.

\begin{figure}
    \centering
    \includegraphics[width=0.8\columnwidth]{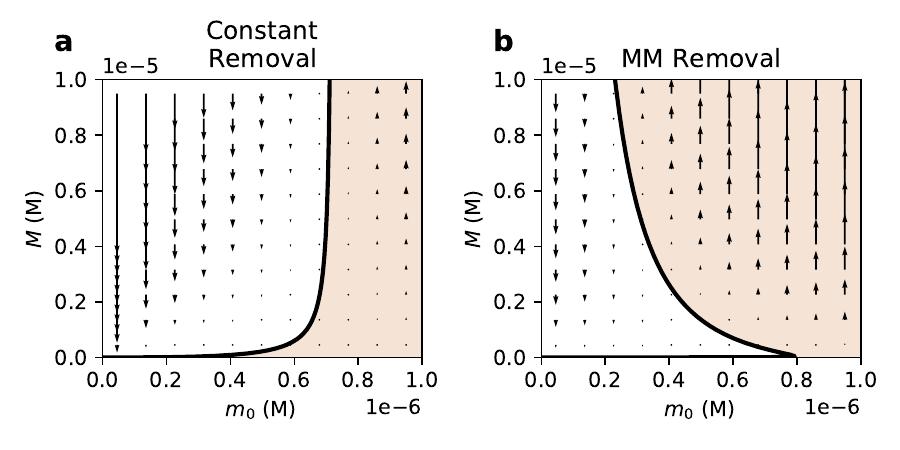}
    \caption{Flow in the $m_0-M$ plane (both expressed in molars M)  from the reduced models for constant and enzyme-like removal kinetics. The background colour shows the direction of the flow.
    (a) Reduced model from (\ref{eq:reducedConstClear}) with $k_n=4800.0\text{M}^{-1}\text{hr}^{-1}$, $\lambda=1.0$, $k_2=1.1\times10^{10}\text{M}^{-2}\text{hr}^{-1}$ and $\bar{l}^*$ from \eqref{eq:lbarConstClear}.
    (b) Reduced model from (\ref{eq:reducedMM}) with $k_n=5000.0\text{M}^{-1}\text{hr}^{-1}$, $\lambda=2.5\text{hr}^{-1}$, $K_{\lambda, P}=1.0\times10^{-8}\text{M}$, $k_2=4.0\times10^{10}\text{M}^{-2}\text{hr}^{-1}$ and $\bar{l} = \sqrt{2 \kon/k_2 m_0}$.
    Additional system parameters in (a) and (b) are $n_2=2$, $\nc=2$, $\kon= 1.2\times10^{8}\text{M}^{-1}\text{hr}^{-1}$.}
    \label{fig:flowMM_P}
\end{figure}

When the removal is proportional to the aggregate concentration the fate of the system will be entirely determined by the monomer concentration (Section \ref{subsec:constantclearance}, Fig.~\ref{fig:steadyConstClear}). We can reduce the three variable description ($M$, $P$ and $m_0$) to a two parameter description by prescribing the average length of the aggregates, $\bar{l}$. In the case of unbounded proportional removal, the steady state is given by $\mathbf{q}=\mathbf{A}^{-1}\mathbf{b}$ and so an obvious choice for $\bar{l}$ is to choose the steady state value, $\bar{l}^*=M^*/P^*$, given by (\ref{eq:lbarConstClear}), which exactly recovers the same steady state mass ($M^*$) as the full model. With this assumption, the evolution equation for $M$ becomes
\begin{equation}
    \frac{\text{d}M}{\text{d}t}= \nc k_n m_0^{\nc} + n_2 k_2 m_0^{n_2} M + 2 \kon m_0 M/\bar{l}^* - \lambda M.
    \label{eq:reducedConstClear}
\end{equation}
This system is now one dimensional, but it is useful to see it in the plane $(m_0,M)$ by adding the equation $\text{d}m_0/\text{d}t=0$. This reduced  system captures the key features of the aggregation kinetics: the approach to the steady state for $m_0 < m_0^{\text{(crit)}}$ and runaway aggregation for $m_0 > m_0^{\text{(crit)}}$. A phase plane analysis of this model clearly shows the dynamics and the emergence of the stability as shown in  Fig.\ref{fig:flowMM_P}(a). The average length assumption projects the two dimensional $M-P$ plane flows onto a line in $M-P$ space, shown by the dotted line in Fig.~\ref{fig:steadyConstClear}(b) and (c). The key features of the dynamics are well captured by the projection onto this line.

We can similarly reduce the system to a two-parameter description when the removal is bounded to give
\begin{equation}
    \frac{\text{d}M}{\text{d}t}= \nc k_n m_0^{\nc} + n_2 k_2 m_0^{n_2} M + 2 \kon m_0 M/\bar{l} - \frac{\lambda M}{1+M/\bar{l}K_{\lambda, P}}.
    \label{eq:reducedMM}
\end{equation}
However, the transition to disease occurs at high aggregate mass and so we use the average length for large aggregate mass given by (\ref{eq:lbar-growth}). The fixed points of this reduced system are given by a quadratic equation in $M$ which leads to two steady states when $m_0$ is low and no steady states when $m_0$ is high. This reduced model captures the key features of the full model, with the existence of a stable and unstable fixed point and a critical monomer concentration. However, the exact value at which this bifurcation occurs is slightly perturbed in the reduced model compared to the full model. This is due to the approximation for $\bar{l}$ being accurate only when the aggregate steady state mass is large and, indeed, Fig.~\ref{fig:steadyMM} shows that the two models agree well in this region.

The fate of the system now depends on both the monomer concentration and the aggregate mass. The system is only stable, and therefore healthy, if both the monomer concentration and aggregate mass are low. Intuitively this dependence on the aggregate mass makes sense. If there are more aggregates (number and mass) in the system, then the rates of secondary nucleation and elongation will increase. When the aggregate removal rate is high enough to outcompete these effects, the aggregate mass/number is reduced and approaches a steady state value where the production and removal are balanced. For very large aggregate mass/number, the production rates of aggregates continue to increase with increasing aggregate mass/number, however for physically realistic models of removal mechanisms the removal rate of aggregates will begin to saturate and at some point will no longer be able to balance the aggregate production/elongation rates, leading to runaway aggregation.

These reduced dynamics show the different mechanisms through which a system can transition from a healthy state into a pathological state. A stable steady state could transition to runaway aggregation if the monomer in the system was increased beyond the critical value. Alternatively, the sudden introduction of aggregate mass would cause the system to move upwards on the phase plane and cross into the unstable regime. This transition can only occur when a system allows for  both stable or unstable states at the same monomer concentration. Such a system could also transition from an unstable to a stable state by removing aggregate mass from the system (moving down on the phase plane). This is likely to be hard to achieve therapeutically. However changes in monomer concentration, moving the left on the phase plane and hopefully into the stable region, can be achieved through monomer reducing therapies~\cite{cole_-synuclein_2021}. How much of an intervention is required to cause a system to change state depends on the aggregate mass in the system and so this suggests that such therapies would only be effective at early time after the transition to disease, when the aggregate mass is still small. This reduced parameter system provides a useful perspective to discuss the landscape of NDDs and captures the effects of seeding and monomer reduction therapies.

\section{Phase-plane structure is preserved across aggregation and removal kinetics}
\label{sec:phaseplanestructure}

When developing the model of bounded removal in Section \ref{subsec:boundedclearance} we focused on the kinetics of an enzyme-like mediated removal mechanism with binding rates independent of aggregate size. Crucially, this model predicts an upper unstable branch in the phase plane of the $M-m_0$ dynamics, i.e. that the introduction of seeds can push the system to runaway aggregation even below the critical monomer concentration. 

This phenomenon is generic and found in a larger class of aggregation and removal mechanisms that all share the same bifurcation structure. To demonstrate the wide scope of this theory, we show that this bifurcation structure is preserved for a range of different removal mechanisms, with different functional dependence, and for a variety of aggregation kinetics. We overview these different mechanisms below and plot the resulting phase planes in Fig \ref{fig:generalReduced}. Table \ref{tab:generalised} summarises the different models.

\subsection{Generalised models of aggregate removal}
\label{sec:genClear}

\subsubsection{Enzyme-like kinetics saturating in aggregate mass}\label{sec:M-saturating}

In Section \ref{subsec:boundedclearance} we assumed that the binding rate, $k_i^{b}$, dissociation rate, $k_i^{d}$ and breakdown rate $k_i^{c}$ were constant. With these assumptions, the net aggregate removal followed MM kinetics, with the rate saturating for high aggregate number concentration, $P$. Different assumptions can be made about the length dependence of the rates, we now showcase that the characteristic shape of the phase plane is maintained also under different assumptions. Specifically, when ${k_i^b}/{k_i^d} \propto i$ and $k_i^c \propto i^{-1}$, we again recover MM kinetics. However, in this case, the rate saturates with respect to the aggregate mass concentration, $M$. This scaling is inspired by a binding-and-removal model, where a removal component is more likely to bind to larger aggregates. However, the monomer-monomer bonds are cleaved at a constant rate, and the time to remove the whole aggregate is thus proportional to its size. The removal rates for the number and mass concentrations are now $\lambda P/(K_{\lambda, M}+M)$ and $\lambda M/(K_{\lambda, M}+M)$ respectively. Setting $k_i^b/k_i^d = \beta i$ and $k_i^c=\gamma i^{-1}$ we find $\lambda = \beta\gamma f_e^T(t)$ and $K_{\lambda, M}=\beta^{-1}$. The reduced model for the system is
\begin{equation}
\frac{\text{d}M}{\text{d}t}= \nc k_n m_0^{\nc} + n_2 k_2 m_0^{n_2} M + 2 \kon m_0 M/\bar{l} - \frac{\lambda M}{1+M/K_{\lambda, M}}.
\label{eq:reducdMMinM}
\end{equation}

The reduced dynamics for this model are shown in Fig.~\ref{fig:generalReduced}(a) and maintain the characteristic shape. 

\subsubsection{Multiple removal mechanisms}

\begin{figure}
    \centering
    \includegraphics[width=0.6\columnwidth]{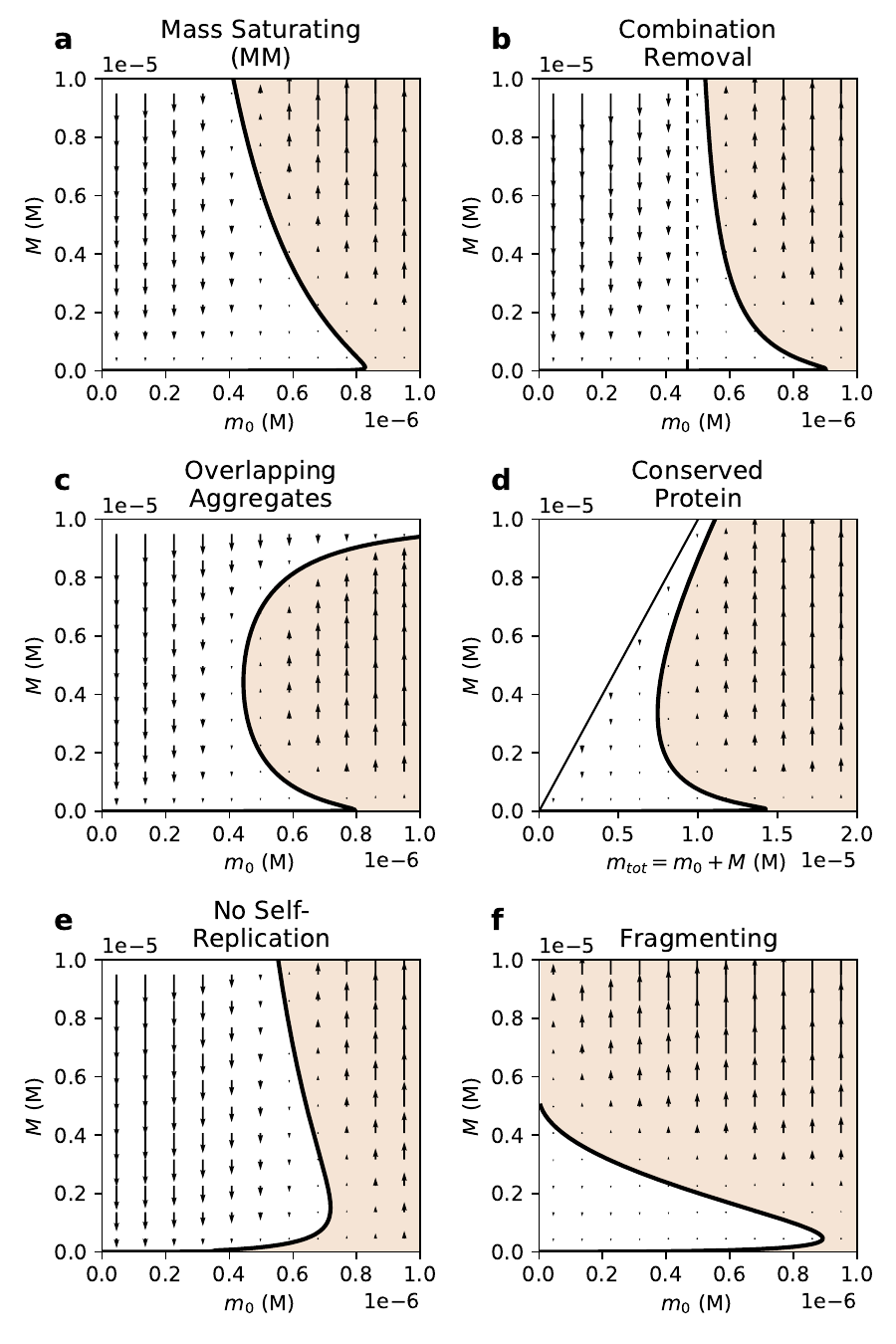}
    \caption{Reduced dynamics for the generalised models. The flows in the different plots are given by (a) (\ref{eq:reducdMMinM}) with $\lambda=2.5\text{hr}^{-1}$, $K_{\lambda, M}=5.0\times10^{-6}\text{M}$
    (b) (\ref{eq:mutliplemechs}) with $\lambda=3\text{hr}^{-1}$, $K_{\lambda, P}=1.0\times10^{-8}\text{M}$, $\alpha=2/3$
    (c) (\ref{eq:overlap}) with $\lambda=2.5\text{hr}^{-1}$, $K_{\lambda, P}=1.0\times10^{-8}\text{M}$, $\rho=1\times10^{5}\text{M}^{-1}$
    (d) (\ref{eq:conserved}) with $\lambda=250\text{hr}^{-1}$, $K_{\lambda, P}=2.0\times10^{-8}\text{M}$.
    Additional system parameters in (a)---(d) are $k_n=5000.0\text{M}^{-1}\text{hr}^{-1}$, $\kon= 1.2\times10^{8}\text{M}^{-1}\text{hr}^{-1}$, $\nc=2$, $n_2=2$, and $k_2=4.0\times10^{10}\text{M}^{-2}\text{hr}^{-1}$. $\bar{l}=\sqrt{2 \kon/k_2 m_0}$ in (a)---(c) and $\bar{l}=\sqrt{2 \kon/k_2 m_{\textrm{tot}}}$ in (d). Flows given by (e) \eqref{eq:noSecNucReduced} with $k_n=5000\text{M}^{-1}\text{hr}^{-1}$, $\kon= 5\times10^{7}\text{M}^{-1}\text{hr}^{-1}$, $\nc=2$, $\lambda=1\text{hr}^{-1}$, $K_{\lambda, M}=1.5\times10^{-6}\text{M}$, $P=\sqrt{M k_n m_0^{\nc-1} /\kon}$ and (f) \eqref{eq:fragmentingReduced} with $k_-=0.0075\text{hr}^{-1}$, $k_n=50000\text{M}^{-1}\text{hr}^{-1}$, $\kon= 2\times10^{8}\text{M}^{-1}\text{hr}^{-1}$, $\nc=2$, $\lambda=2\text{hr}^{-1}$, $K_{\lambda, P}=2.0\times10^{-8}\text{M}$ and $\bar{l}=\sqrt{2 \kon m_0/k_-}$}
    \label{fig:generalReduced}
\end{figure}

We expect that the complex regulation of living systems will result in multiple mechanisms to remove aggregates, each  with potentially different kinetics and saturation behaviour. Again, when the aggregation rates outpace the removal mechanisms we recover the same bifurcation structure. Consider for instance, the case of a system combining a removal proportional to mass and a MM-like removal that saturates in aggregate number (the removal mechanisms in (\ref{eq:reducedConstClear}) and (\ref{eq:reducedMM})). At low aggregate mass, the rate is proportional, with constant $\lambda$.  We denote the relative contribution of the saturating mechanism with $\alpha$. The combination of these two removal mechanisms means that we find that the bifurcation point moves to a higher monomer concentration compared to each removal mechanism separately, but that the bifurcation structure remains unchanged, as can be seen in Fig.~\ref{fig:generalReduced}(b). At low monomer concentration, we have global stability due to constant removal, which for higher monomer concentrations is outpaced by aggregation and the MM-like removal gives the familiar stability structure. The reduced model for this system is
\begin{equation}
\begin{split}
\frac{\text{d}M}{\text{d}t}&= \nc k_n m_0^{\nc} + n_2 k_2 m_0^{n_2} M + 2 \kon m_0 M/\bar{l}\\
&-\lambda M\left((1-\alpha) + \frac{\alpha}{1+M/\bar{l}K_{\lambda, P}}\right).
\end{split}
\label{eq:mutliplemechs}
\end{equation}
This combination model is shown in Fig.~\ref{fig:generalReduced}b and again displays the same characteristic phase plane.

\begin{table}[]
    \centering
    {\fontsize{10}{8}\selectfont
    \begin{tabular}{|c|c|c|}
    \hline
    \textbf{Model Name} & \textbf{Reduced }$\mathbf{\textbf{d}M/\textbf{d}t}$ & \textbf{Phase Plane}\\
    \hline
    Constant Removal & $\nc k_n m_0^{\nc} + n_2 k_2 m_0^{n_2} M + 2 \kon m_0 M/\bar{l}^* - \lambda M$ & Fig.\ref{fig:flowMM_P}(a) \\
    MM Removal & $\nc k_n m_0^{\nc} + n_2 k_2 m_0^{n_2} M + 2 \kon m_0 M/\bar{l} - \frac{\lambda M}{1+M/\bar{l}K_{\lambda, P}}$ & Fig.\ref{fig:flowMM_P}(b) \\
    Mass Saturating (MM)  & $\nc k_n m_0^{\nc} + n_2 k_2 m_0^{n_2} M + 2 \kon m_0 M/\bar{l} - \frac{\lambda M}{1+M/K_{\lambda, M}}$ & Fig.\ref{fig:generalReduced}(a) \\
    Combination Removal & $\nc k_n m_0^{\nc} + n_2 k_2 m_0^{n_2} M + 2 \kon m_0 M/\bar{l}-\lambda M\left((1-\alpha) + \frac{\alpha}{1+M/\bar{l}K_{\lambda, P}}\right)$ & Fig.\ref{fig:generalReduced}(b) \\
    Overlapping Aggregates  & $\nc k_n m_0^{\nc} + n_2 k_2 m_0^{n_2} (M- \rho M^2)+ 2 \kon m_0 (M- \rho M^2)/\bar{l}- \frac{\lambda M}{1+M/\bar{l}K_{\lambda, P}}$ & Fig.\ref{fig:generalReduced}(c) \\
    Conserved Protein  & $\nc k_n (m_{tot}-M)^{\nc} + n_2 k_2 (m_{tot}-M)^{n_2} M+ 2 \kon (m_{tot}-M) M/\bar{l} - \frac{\lambda M}{1+M/\bar{l}K_{\lambda, P}}$ & Fig.\ref{fig:generalReduced}(d) \\
    No Self-Replication  & $\nc k_n m_0^{\nc} + 2 \kon m_0 P(M) - \frac{\lambda M}{1+M/K_{\lambda, M}}$ & Fig.\ref{fig:generalReduced}(e) \\
    Fragmenting  & $\nc k_n m_0^{\nc} + 2 \kon m_0 M/\bar{l} - \frac{\lambda M}{1+M/\bar{l}K_{\lambda, P}}$ & Fig.\ref{fig:generalReduced}(f) \\
    \hline
    \end{tabular}}
    \caption{Summary of different systems across aggregation and removal kinetics.}
    \label{tab:generalised}
\end{table}

\subsection{Generalised models of aggregation}
\label{sec:genAgg}

\subsubsection{Finite disease steady state from aggregate \textit{clumping}}

When a cell is in the pathological state the aggregate mass grows exponentially. However for a cell of finite volume and limited resources this growth can only take place for finite time. Autopsy data show that diseased cells are often full of aggregates (for example see coronal and sagittal planes~\cite{iba_synthetic_2013} and aggregate renders~\cite{guo_situ_2018}) and so at some point the aggregate mass approaches a cellular carrying capacity. One of the potential contributors to this carrying capacity is the fact that, as the aggregate concentration increases, multiple aggregates may touch or overlap and disrupt the elongation and self replication processes. We now focus on this process as an example, to show the key parts of the bifurcation structure are preserved. If we assume that each aggregate is independently distributed within the volume of the cell, then we expect the probability of two aggregates being separated by a distance less than some threshold to increase as $M^2$ and so the aggregation kinetics with this correction to account for inactive aggregate surface becomes
\begin{equation}
\begin{split}
\frac{\text{d}M}{\text{d}t}=& \nc k_n m_0^{\nc} + n_2 k_2 m_0^{n_2} (M- \rho M^2)\\
&+ 2 \kon m_0 (M- \rho M^2)/\bar{l}- \frac{\lambda M}{1+M/\bar{l}K_{\lambda, P}}
\end{split}
\label{eq:overlap}
\end{equation}
where $\rho$ is the overlap constant. Fig.~\ref{fig:generalReduced}c shows the reduced dynamics for this system which appear similar to the ones found before, however there is now a pathological stable branch at high aggregate concentrations. For very low monomer concentration there is only one steady state as the removal mechanisms can keep up with the aggregation and the steady state is determined by the balance of aggregation and removal. For intermediate monomer concentrations, there is still a stable steady state at low aggregate concentrations, however, for larger aggregate mass the crowding effects slow the rate of aggregation, resulting in a second stable steady state. Above the critical monomer concentration, there is only a high aggregate mass, pathological steady state, which is globally attracting. We interpret this exactly as before, however now the aggregate mass in the pathological state is finite at some high aggregate mass, which is more consistent with finite-size cells and parallels what is seen in human pathology. The kinetic model in \eqref{eq:overlap} achieves a bound on aggregate mass by considering a reduced aggregate surface due to overlaps, however other mechanisms, such as a reduction in the monomer production rate, may reduce the aggregation rate and similarly result in a maximum aggregate mass. The exact molecular mechanisms do not affect the observed macroscopic phenomena, but might slightly alter the concentrations at which transitions between the system states occur.

An special case to note here is the situation when the carrying capacity is below the concentration at which removal mechanisms become overwhelmed. In that case the seeding effect will disappear, i.e. the low aggregate stable branch smoothly becomes the high aggregate mass stable state, with no bifurcation, and for all monomer concentrations there is only one steady state in the system. In the above example of clumping, this happens when critical point from the balance of removal and aggregation (the lower turning point on the `S' curve) would occur above the saturating effects due to clumping, $M\approx\rho^{-1}$. We can approximate the aggregate mass at which these effects occur (see See Appendix \ref{app:noseeding}) and derive the following condition for the seeding transition to occur
\begin{equation}
    \rho^{-1}>\frac{2 k_{+} K_{\lambda, P}}{\lambda}\left(\frac{\lambda^2}{2 k_2 k_{+}}\right)^{\frac{1}{1 + n2}}.
\end{equation}

\subsubsection{Constant total protein}

The models of aggregation considered so far assume constant free monomer concentration in the cell, motivated by cellular homeostasis regulating this concentration. The opposite limiting behaviour is to assume the aggregation kinetics are much faster than the expression of the monomer and so the total protein in the system, $m_{tot} = M + m$ is constant. Most real cellular systems will lie on the spectrum between these two extremes. Assuming the MM-like removal mechanism, the rate of change of aggregate mass is still given by (\ref{eq:reducedMM}) and substituting $m_0=m_{tot}-M$, so that $\dot{M}=-\dot{m}$. This corresponds to removal mechanisms converting the aggregated mass back into monomer to conserve total protein. The dynamics of the new reduced model can be seen in a phase plane of $M$ and $m_{tot}$ shown in the Fig.~\ref{fig:generalReduced}(d). The equation describing the system is
\begin{equation}
\begin{split}
\frac{\text{d}M}{\text{d}t}&= \nc k_n (m_{tot}-M)^{\nc} + n_2 k_2 (m_{tot}-M)^{n_2} M\\
&+ 2 \kon (m_{tot}-M) M/\bar{l} - \frac{\lambda M}{1+M/\bar{l}K_{\lambda, P}},
\end{split}
\label{eq:conserved}
\end{equation}
where $\bar{l}$ is the average aggregate length assumed to be established early in the aggregation dynamics, so that $\bar{l}=\sqrt{2k_{+} / k_2 m_{tot}}$.

Since $\dot{M}=-\dot{m}$, the effects of the conserved total protein will be seen when the aggregate mass concentration is of the same order of magnitude as the monomer concentration. We choose rate parameters o demonstrate this coupling and show a wide range of total aggregate mass. The, by definition, unfeasible region where $M>m_{\textrm{tot}}$ is included but left blank. From Fig.~\ref{fig:generalReduced}(d) it can be seen that system dynamics for conserved total protein has similar structure to the Fig.~\ref{fig:generalReduced}(c), except with an increasing upper branch. The upper branch for large monomer concentration has almost all protein in the aggregated state. Seeding this system with a mass concentration of seeds, $M_{\textrm{seed}}$ increases both the mass concentration $M\rightarrow M+M_{\textrm{seed}}$ and the total protein concentration, $m_{\textrm{tot}}\rightarrow m_{\textrm{tot}}+M_{\textrm{seed}}$ and so corresponds to a diagonal upwards-right move on the phase plane.

This model, (\ref{eq:conserved}), and the model in (\ref{eq:reducedMM}) represent two distinct regimes: the aggregation kinetics are much faster than the homeostatic mechanisms or the inverse. The full system is likely to couple these timescales. In this case,the arrows in the phase planes will be tilted rather than vertical. It is harder to determine the fate of the system from the coupled dynamics, but by considering the two limiting regimes we recover the same macroscopic behaviour and features of the phase plane, indicating that they will also be retained at intermediate points.

\subsubsection{Systems without self-replication}
\label{sec:noselfrep}

We also consider the case of aggregating systems with no self-replication~\cite{meisl_uncovering_2022}, corresponding to $k_2=0$. In these systems, there also exist parameter regimes that exhibit the same phenomenology as the models discussed above. However, the seeding behaviour is only observed when the system saturates with increasing mass concentration, rather than saturating by increasing aggregate number concentration. Simply setting $k_2=0$ and using the number saturating kinetics, given by equation \eqref{eq:dP_MM_P} we find that when $\lambda > k_n m_0^{\nc}$ their exists a finite stationary aggregate number and mass concentration given by
\begin{equation}
    P^* = \frac{K_{\lambda, P} k_n m_0^{\nc}}{\lambda - k_n m_0^{\nc}}
\end{equation}
and
\begin{equation}
    M^* = \frac{ K_{\lambda, P} k_n \left(\kon m_0^{1+\nc} K_{\lambda, P} + 2\lambda m_0^{n_2}- 2 k_n m_0^{\nc+n_2}\right)}{\left(\lambda - k_n m_0^{\nc}\right)^2}.
\end{equation}
This steady state is attracting for all initial values of $M$ and $P$. When $\lambda < k_n m_0^{\nc}$, $M$ and $P$ increase without bound. Phenomenologically, this creates a similar transition from the healthy to the pathological state as observed in the case of unbounded, proportional removal, as there exists a critical monomer concentration that switches the state of the system for all values of the initial aggregate mass.

However, when the removal saturates in $M$ as described in Section \ref{sec:M-saturating} then the model shows seeding susceptibility: the system stability depends on the initial aggregate distribution. The equation describing the reduced model is
\begin{equation}
\frac{\text{d}M}{\text{d}t}= \nc k_n m_0^{\nc} + 2 \kon P(M) - \frac{\lambda M}{1+M/K_{\lambda, M}}
\label{eq:noSecNucReduced}
\end{equation}
where $P$ is a function of $M$, which is needed to reduce the dynamics. In the other reduced models, this was done by assuming some constant average length derived from the large aggregate limit. Here, when there is no self-replication, the length distribution does not reach a steady state and thus does not have a constant average length. We use instead the approximation that $P=\sqrt{M k_n m_0^{\nc-1} /\kon}$ (see appendix \ref{app:noselflength}). Nonetheless, this still produces the same bifurcation structure as before in the reduced monomer-aggregate phase plane (see Fig.~\ref{fig:generalReduced}(e)). Therefore, systems that lack self-replication may not display a seeding effect.

\subsubsection{Systems with fragmentation}

We now consider a system where the only self-replication is due to fragmentation, setting $k_2=0$ and $k_- \neq 0$. The reduced model for this system becomes
\begin{equation}
\frac{\text{d}M}{\text{d}t}= \nc k_n m_0^{\nc} + 2 \kon m_0 M/\bar{l} - \frac{\lambda M}{1+M/\bar{l}K_{\lambda, P}}.
\label{eq:fragmentingReduced}
\end{equation}
The fragmentation rate affects the dynamics by altering the length distribution with the average length being given by $\bar{l}=\sqrt{2 \kon m_0/k_-}$, which similarly to before is a good approximation at large aggregate mass.
Again, these aggregation kinetics give the same bifurcation structure as before, see Fig.~\ref{fig:generalReduced}(f).

\subsection{Sufficient conditions for bistability and seeding transition}

Comparing a range of aggregate production/removal kinetics reveals two sufficient conditions that give rise to the observed disease phenomena. Specifically, we find bistability and the seeding transition in systems that have:
\begin{enumerate}
    \item A self-replicating aggregation mechanism with a monomer concentration-dependent rate; and
    \item A limited capacity aggregate removal mechanism.
\end{enumerate}
Different molecular processes can satisfy these constraints. These two conditions naturally emerge from considering the universal features of aggregate formation \cite{meisl_uncovering_2022} and the physical constraints on the aggregate removal mechanisms. Additionally, other regimes, for example a system with no self-replication, as in Section \ref{sec:noselfrep}, can still show the same stability structure. Moreover, feedback from, for example, aggregate toxicity or damage to vasculature~\cite{ahern_modelling_2025} can limit the capacity of aggregate removal mechanisms, similar to the removal mechanisms that require an intermediary clearance component, such as the Michealis-Menten mechanisms discussed above.

\begin{figure}
    \centering
    \includegraphics[width=0.95\columnwidth]{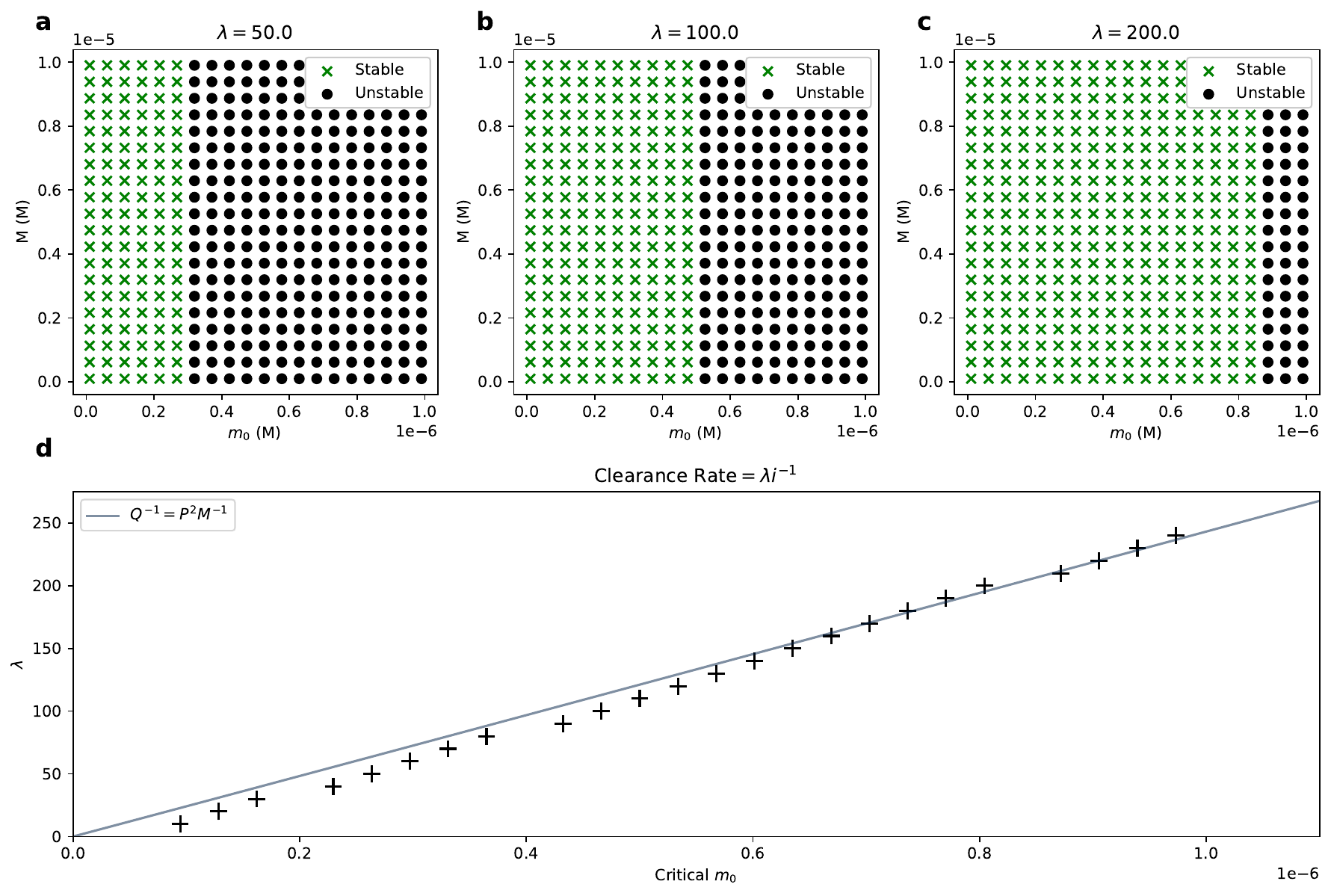}
    \caption{Comparison of the stability of aggregating systems with length-dependent removal, given by (\ref{eq:pi_len_dep}). (a)-(c) Summarise the results from numeric simulation for different removal rates, and shows the stability boundary occurs at a critical monomer concentration with no seeding transition (each points is a unique simulation). (d) The critical monomer concentration at which runaway aggregation occurs is shown for different $\lambda$ for both the simulation results and the analytic approximation. The initial aggregate mass simulated is $M=5\times10^{-5}\text{M}$. The other system parameters are $k_n=4800\text{M}^{-1}\text{hr}^{-1}$, $\kon= 1.2\times10^{8}\text{M}^{-1}\text{hr}^{-1}$, $\nc=2$, $n_2=2$, $k_2=1.1\times10^{10}\text{M}^{-2}\text{hr}^{-1}$, and $\nu=-1$. The threshold to determine if the aggregation was bounded is $1\times10^{-3}\text{M}$. The simulations used $N=10000$, $dt=0.001\text{hr}$, and $T=1000\text{hr}$.}
    \label{fig:numinus1}
\end{figure}

\section{Length-dependent removal alone does not yield cellular bistability}

Having discussed the sufficient conditions for bistability and seeding behaviour above, we now briefly investigate the degree to which these are also necessary conditions. We have already established that even without self-replication, some systems can show bistability. Another obvious modification to the above models is to allow the removal rates to depend on the size of the aggregates. To illustrate this, we first consider a situation where the rate of removal is reduced for longer aggregates~\cite{fertanClearanceBetaamyloidTau2025}. This is also a physically meaningful regime, as longer aggregates may take longer to clear from living cells and tissue as there are more bonds to break. It is useful to understand if this extra mechanism will capture the observed seeding behaviour even with an unbounded removal process; that is, whether a sudden increase in the concentration of aggregates, including longer and more persistent fibrils, would cause a transition to runaway aggregation even when removal processes have no maximum capacity. Longer aggregates take more time to clear, so one might suppose that by the time they are removed from the system, the nucleation and subsequent elongation will have already replenished and then increased the concentration of longer aggregates, leading to positive feedback. However, a detailed mathematical description shows this is not the case.

We expect the aggregate removal rate to be inversely proportional to the number of bonds, which in turn is proportional to aggregate length. We assume $\lambda_i = \lambda i^{\nu} f(t, i)$, where we choose $\nu=-1$. For the non-fragmenting, non-depolymerising system discussed here ($k_{-}=0$, $k_\text{off}=0$), the master equation for this model is
\begin{equation}
    \begin{split}
    \frac{\text{d}f(t, i)}{\text{d}t} &= \delta_{i, n_C}k_n m^{n_C} + \delta_{i, n_2} k_2 m^{n_2} \left(\sum_{j=\nc}^{\infty} j f(t, j)\right)\\
    &+ 2\kon m (f(t, i-1)-f(t, i)) - \lambda i^{-1} f(t, i).
    \end{split}
    \label{eq:pi_len_dep}
\end{equation}
Unlike the previous removal models, the evolution of the aggregate number, $P$, depends on the $-1^\text{th}$ moment of the length distribution, $Q^{-1}=\sum_i i^{-1} f(t, i)$. This does not result in a closed system of moment equations that can be used to determine the transition to disease. Instead, we can numerically evolve the system for different initial values of $M$ and $m_0$ and observe the resulting dynamics. As described in Appendix \ref{app:numdeets} we simulate each aggregate population using the master equation and a fourth-order Runge-Kutta method. When the final aggregate mass, $M$, is greater than a threshold value, we label the system as unstable, corresponding to a pathological state. Panel (a)-(c) of Fig.~\ref{fig:numinus1} reports the results of these simulations for three different removal rates, $\lambda$. It is interesting to note that this removal mechanism does not exhibit a seeding-like transition. The stability of a state does not depend on the initial aggregate mass concentration, but exclusively on the monomer concentration, similar to the case of a length independent constant removal rate (Section \ref{subsec:constantclearance}).

This effect can be understood mathematically. Since the evolution of all the $f(t, i)$s defines an infinite system of linear coupled equations, it only permits one steady state which describes the stationary length distribution. If this steady state is positive then it will be attractive for all positive concentrations. If it is not positive, then the aggregate mass concentration will increase without bound. This argument is general for any proportional removal regardless of the length dependence, $\lambda_i=g(i) f(i, t)$, where $g(i)$ is some length-dependent function.    

For $\nu=-1$ the moment equations are not closed, specifically because the evolution of $P$ depends on $Q^{-1}$. For a sharply peaked distribution we can approximate the $Q^{-1}$ moment by setting $Q^{-1}=P^2/M$, equivalent to assuming all aggregates have the same length. The moment equations then become
\begin{align}
    \frac{\text{d}M}{\text{d}t}&= \nc k_n m_0^{\nc} + n_2 k_2 m_0^{n_2} M + 2 \kon m_0 P - \lambda P
    \label{eq:nudependentM} \\
    \frac{\text{d}P}{\text{d}t}&= k_n m_0^{\nc} + k_2 m_0^{n_2} M - \lambda \frac{P^2}{M}.
    \label{eq:nudependentP}
\end{align}
These equations permit stationary states, $M^*$ and $P^*$, which are positive and real when the system is stable. For the specific case of $\nc=n_2=2$, we find that the system is only stable when $\lambda > 2 \kon m_0$. Panel (d) of Fig.~\ref{fig:numinus1} demonstrates that this prediction of the critical monomer concentration is a good approximation to the numerical simulation. This model further generalises the necessity of a removal mechanism with limited capacity to describe disease relevant phenomena, even for length-dependent removal mechanisms.

\section{Conclusions}

This work extends the existing \invitro{} models of pathological protein aggregation to include physically realistic removal processes that occur \invivo{} to describe the aggregation state of cells in living systems. Crucially we showed that typical disease features can be explained when the \invivo{} aggregate removal mechanisms have a limited capacity. This is a very natural physical constraint, and we showed that it holds for a range of specific aggregation and removal mechanisms. The transition to disease via seeding is only predicted when the removal has this limited capacity - even if the removal mechanisms depend on other system parameters such as the aggregate length distribution.

This description of a balance between aggregation and active removal complements the simpler nucleation-limited descriptions, where there is no significant removal but the production of the initial aggregate via primary nucleation is so slow that spontaneous formation essentially never occurs. Cellular systems in this nucleation-limited regime will behave in the same way as non-living compartmentalised systems, such as the nucleation of purified protein in micro-droplets~\cite{Knowles2011a,Michaels2018}: the dynamics are governed by the stochastic nucleation of random cells~\cite{Sinnige2021} and are extremely susceptible to the introduction of seeds~\cite{pfammatter_absolute_2017}.  Prion diseases are a real life example of such a system. By contrast, in the more common aggregation-related diseases, such as Parkinson's and Alzheimer's disease, our descriptions are required as removal appears to play a central role, as evidenced by the fact (1) low aggregate amounts are present even in health~\cite{andrewsLargescaleVisualisationAsynuclein2024,boken_singlemolecule_2024}, (2) decline in removal mechanism with ageing is a risk factor~\cite{Labbadia2015, hardy_genetics_2025} and (3) aggregates disappear when the aggregation rates are decreased~\cite{mummery_tau-targeting_2023, cole_-synuclein_2021}.
\begin{figure}
    \centering
    \includegraphics[width=0.7\columnwidth]{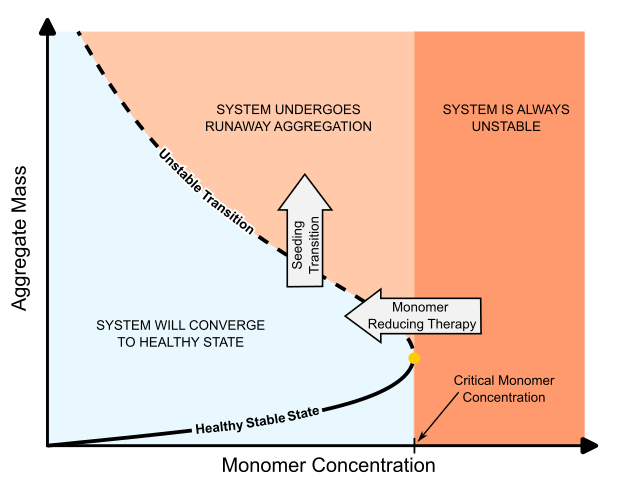}
    \caption{Schematic of the typical monomer-aggregate phase plane structure highlighting the key features that govern transitions to disease.}
    \label{fig:gentrans}
\end{figure}
To provide a convenient representation of cellular aggregation dynamics, we have developed a visual representation of  monomer-aggregate mass phase planes. Fig.~\ref{fig:gentrans} is a scheme showing the characteristic phase plane structure. The stability of a cell is naturally reflected by the fixed points in this space and these two parameters can capture many relevant disease associated phenomena and effects of therapies. As such, this phase plane is a convenient \textit{map} to guide dosing strategies or future therapeutics and we believe it can serve as an easily accessible, central tool for understanding aggregation in disease. Additionally, working with this reduced dimension system will be useful in ongoing efforts to connect molecular mechanisms to tissue-wide spread of disease~\cite{huangCelllevelModelPredict2025, huangTauAccumulationPatterns2024}, without performing full simulation of the entire master equation.

Understanding the magnitude of the competing aggregation and removal processes \invivo{} is essential to understanding disease emergence and progression and thus designing effective treatment strategies. Our model presents a unified theory that can describe different disease phenomena within the same mechanistic framework. As experimental data increases in resolution and abundance, this theory will enable quantification of the different parameters, allowing comparison across diseases and predictions of the effectiveness of therapeutic interventions. 

\appendix

\section{Numerical Details}\label{app:numdeets}

Moving from the master equation to the moment description of the kinetics, we have assumed the system can support infinite aggregates and, for depolymerising systems, that $\nc f(t, \nc) \ll M$. Numerical simulation of the dynamics of the system at the level of the master equation can verify that this coarse grained description provides an accurate summary and prediction of the kinetics. Simulating an infinitely large aggregate presents computational challenges, and so here, we introduce a maximum aggregate length that does not grow by elongation but is still removed via the removal mechanisms~\cite{thompson_role_2021}. A physical justification behind this assumption is the fact that cells have a finite size and so cannot support infinitely long aggregates. We use similar values to those from \invitro{} measurements of A$\beta$40~\cite{cohen_proliferation_2013} and A$\beta$42~\cite{Meisl2014, Linse2020}.

We simulate the system by explicitly evolving the aggregate population at each length $i$ for $i$ from $\nc$ to $N+\nc$, where $N$ is a simulation parameter. We prescribe some initial length distribution at time $t=0$ and the system is evolved using a fourth order Runge Kutta method that explicitly updates the aggregate population at each length at time intervals of $\text{d}t$ from $t=0$ until $t=T$, where again $\text{d}t$, $T$ and the initial length distribution are simulation parameters. The theoretical predictions describe systems with a maximum length significantly larger than the mean length distribution and we can additional ensure that any differences between the full and truncated systems are small by ensuring that $\alpha^N \ll 1$. With the values shown in Fig.~\ref{fig:steadyConstClear} and for $m_0=1\times 10^{-6}\text{M}$ we find that $\alpha^N = 8.7 \times 10^{-19}$ and so we expect the truncated system to give good agreement with the theoretical full infinite system.

\section{Effect of Nucleation Processes on Size Distribution}\label{app:nucpro}

The effect of secondary nucleation on the average length initially seems confusing, and in particular it might seem strange that the dependence on $k_2$ vanishes when $\nc=n_2$, so we briefly expand on this point here. The mathematical form obtained can be explained as aggregates are only nucleated at length $\nc$ or $n_2$ and the population of aggregates at other lengths decays geometrically away from these source terms. When $\nc = n_2$ there is only one source term, therefore the average aggregate length is determined entirely by the rate of decay of the aggregate population with increasing length. If the nucleation rate were increased, the population at length $n_c$ would increase, and subsequently the population of aggregates of every length would increase proportionally, however the decay length would remain the same and the \textit{average} aggregate length would also be unchanged. For $n_2>\nc$ the source term at $n_2$ increases the aggregate population at large lengths and so increases the average length.

\section{Condition for seeding transition in a system with overlapping aggregates.}\label{app:noseeding}

We consider the conditions necessary for the seeding transition to occur in a system with overlapping aggregates (equation \ref{eq:overlap}) to derive an order of magnitude estimate for the presence of seeding. The transition occurs when the characteristic concentration at which the clearance mechanisms become overwhelmed (lower bend of the `S') is lower than the characteristic concentration of aggregation saturation (the upper bend of the `S'). The characteristic concentration of aggregation saturation is given by
\begin{equation}
    M^{\text{upper}}\approx\rho^{-1}.
\end{equation} The aggregate mass at which the clearance mechanisms start to saturate and thus become overwhelmed is given by
\begin{equation}
    M^{\text{lower}}\approx K_{\lambda, P}\bar{l}.
\end{equation}
The lower bend occurs at the critical monomer concentration, given by $\lambda M \approx 2k_+m_0 M/\bar{l}$ (assuming that elongation is the major driver of aggregate mass increase). Combining this with the approximate solution for aggregate length $\bar{l}=\sqrt{2 k_+ /(k_2 m_0^{n_2-1})}$, we find that at $m_0^{\text{(crit)}}$ the average aggregate length is
\begin{equation}
    \bar{l}^{\text{ lower}} = \frac{2 k_{+}}{\lambda}\left(\frac{\lambda^2}{2 k_2 k_{+}}\right)^{\frac{1}{1 + n2}}
\end{equation}

Thus, we only get the two separate regimes, and the S-shape, when $M^{\text{upper}}>M^{\text{lower}}$. This gives the condition for the seeding transition in this system as
\begin{equation}
    \rho^{-1}>\frac{2 K_{\lambda, P} k_{+}}{\lambda}\left(\frac{\lambda^2}{2 k_2 k_{+}}\right)^{\frac{1}{1 + n2}}.
\end{equation}
Important to note is that the left hand side of this inequality is exclusively determined by the typical scale of the cellular carrying capacity so can be easily modified to other kinetic models.

\section{Length distribution with no self-replication}\label{app:noselflength}

We consider a system in which there is no self-replication, new aggregates are produced only from the primary nucleation step and are removed according Michaelis-Menten kinetics that saturate in the aggregate mass. The moment description is
\begin{align}
    \frac{\text{d}M}{\text{d}t}&= \nc k_n m_0^{\nc} + 2 \kon m_0 P - \frac{\lambda M}{1+M/K_{\lambda, M}} \\
    \frac{\text{d}P}{\text{d}t}&= k_n m_0^{\nc} - \frac{\lambda P}{1+M/K_{\lambda, M}}.
\end{align}
When the system is unstable, either due to system parameters or the initial aggregate concentration, then the aggregation will grow without bound. When $M$, $P$ and the average aggregate length are very large the moment dynamics become
\begin{align}
    \frac{\text{d}M}{\text{d}t}&\approx 2 \kon m_0 P\\
    \frac{\text{d}P}{\text{d}t}&\approx k_n m_0^{\nc}
\end{align}
which, after a long time, gives
\begin{align}
    M &\approx \kon k_n m_0^{\nc+1} t^2\\
    P &\approx k_n m_0^{\nc} t
\end{align}
and thus $P\approx\sqrt{Mk_n m_0^{\nc-1}/\kon}$.

\bibliography{phaseplane_refs}

\end{document}